# Insulator-to-Metal Transition and Isotropic Gigantic Magnetoresistance in Layered Magnetic Semiconductors

Gokul Acharya[1], Bimal Neupane[2], Chia-Hsiu Hsu[3,4] Xian P. Yang[5], David Graf[6], Eun Sang Choi[6], Krishna Pandey[7], Md Rafique Un Nabi[1,8], Santosh Karki Chhetri[1], Rabindra Basnet[1,9], Sumaya Rahman[1], Jian Wang[10], Zhengxin Hu[11,12], Bo Da[11], Hugh Churchill[1,7,8], Guoqing Chang[3], M. Zahid Hasan[5,13,14], Yuanxi Wang[2*], Jin Hu[1,7,8*]

[1]Department of Physics, University of Arkansas, Fayetteville, AR, USA

[2]Department of Physics, University of North Texas, Denton, TX, USA

[3]Division of Physics and Applied Physics, School of Physical and Mathematical Sciences, Nanyang Technological University, Singapore, Singapore

[4]Quantum Materials Science Unit, Okinawa Institute of Science and Technology (OIST) Okinawa 904-0495, Japan

[5]Laboratory for Topological Quantum Matter and Advanced Spectroscopy (B7), Department of Physics, Princeton University, Princeton, NJ, USA.

[6]National High Magnetic Field Lab, Tallahassee, FL, USA

[7]Materials Science and Engineering Program, Institute for Nanoscience and Engineering, University of Arkansas, Fayetteville, AR, USA

[8]MonArk NSF Quantum Foundry, University of Arkansas, Fayetteville, Arkansas, USA

[9]Department of Chemistry & Physics, University of Arkansas at Pine Bluff, Pine Bluff, Arkansas 71603, USA

[10]Department of Chemistry and Biochemistry, Wichita State University, Wichita, KS, USA

[11]Center for Basic Research on Materials, National Institute for Materials Science (NIMS), Tsukuba-city, Ibaraki, Japan

[12]Department of Chemistry, University of Cambridge, Cambridge, United Kingdom

[13]Princeton Institute for Science and Technology of Materials, Princeton University, Princeton, NJ, USA.

[14]Lawrence Berkeley National Laboratory, Berkeley, CA, USA.

[*]Correspondence: yuanxi.wang@unt.edu; jinhu@uark.edu

**Magnetotransport, the response of electrical conduction to external magnetic field, acts as an important tool to reveal fundamental concepts behind exotic phenomena and plays a key role in enabling spintronic applications. Magnetotransport is generally sensitive to magnetic field orientations. In contrast, efficient and isotropic modulation of electronic transport, which is useful in technology applications such as omnidirectional sensing, is rarely seen, especially for pristine crystals. Here we propose a strategy to realize extremely strong modulation of electron conduction by magnetic field which is independent of field direction. GdPS, a layered antiferromagnetic semiconductor with resistivity anisotropies, supports a field-driven insulator-to-metal transition with a paradoxically *isotropic* gigantic negative magnetoresistance insensitive to magnetic field orientations. This isotropic magnetoresistance originates from the combined effects of a near-zero spin-orbit coupling of $Gd^{3+}$-based half-filling *f*-electron system and the strong on-site *f-d* exchange coupling in Gd atoms. Our results not only provide a novel material system with extraordinary magnetotransport that offers a missing block for antiferromagnet-based ultrafast and efficient spintronic devices, but also demonstrate the key ingredients for designing magnetic materials with desired transport properties for advanced functionalities.**

Electronic transport is one of the most widely adopted techniques to uncover physical properties of materials. Owing to its sensitivity to density of states and carrier scatterings, applying a magnetic field modulates electronic transport via various mechanisms that are directly associated with the fundamental properties of materials. The competition of diverse contributing mechanisms leads the electrical resistivity to change with the direction and strength of the applied magnetic field. A variety of magnetotransport phenomena have been well-explored, such as weak (anti)localization[1], quantum oscillations[2], giant magnetoresistance (GMR)[3], and colossal magnetoresistance (CMR)[4]. This venerable subject remains rather vibrant in contemporary condensed matter and materials physics research, not only because it provides an effective approach to understand the electronic properties of materials, but also owing to emergence of novel phenomena, such as chiral magnetic effect[5–7] and extremely large positive magnetoresistance (MR)[6–9] in topological quantum materials.

Efficient modulation of electron conduction by magnetic field, which is manifested into resistance increase or decrease and characterized by MR, is scientifically interesting and technologically useful. Generally, MR is sensitive to the direction of the magnetic field with respect to the applied current. Examples include ordinary orbital MR due to Lorentz force that is sensitive to the angle between field and current[10], GMR caused by spin dependent scattering in magnetic multilayers and hence naturally anisotropic[11], and chiral magnetic effects induced by the magnetic field component parallel to the current orientation[5–7]. In contrast, strong MR independent of magnetic field direction is much less common. Ferromagnetic metals and alloys show small MR anisotropy of a few percent, but the amplitude of MR is too small to be technological useful[12]. The traditional CMR effect in correlated materials is considered to be more isotropic than GMR effects, but it generally varies by at least 30% percent depending on magnetic field direction[13,14]. Recently, a few non-oxide materials with strong negative MR have been discovered, such as $EuMnSb_2$[15,16], $EuTe_2$[17,18], $Mn_3Si_2Te_6$[19,20], and $EuCd_2P_2$[21]. These materials all display substantial field-direction dependent MR.

Magnetotransport phenomena typically arise from modulations in carrier density and/or scattering by the applied magnetic field. Therefore, the dominance of anisotropic magnetotransport is not surprising, given that both the magnetic field and electron transport are directional. Anisotropic and strong MR has already found applications in electronic compasses, position and motion sensing, and hard disk drives. On the other hand, isotropic MR, if realized, can provide omnidirectional sensing capabilities, serving as a valuable complement to anisotropic MR-based devices. This synergy allows for the acquisition of comprehensive information for magnetic field sensing and magnetic imaging. Here we propose a strategy to create isotropic colossal MR independent of magnetic field direction in compounds with vanishing magnetic anisotropy, through an insulator-to-metal transition driven by the same-atom *d-f* exchange splitting. We present a magnetic semiconductor GdPS as such a case, in which the zero orbital angular momentum for *f*-electrons yields a near-zero magnetic anisotropy. The substantial magnetization from half-filled *f*-orbitals strongly splits the Gd *d*-bands through an on-site exchange interaction, and results in a metallic state with well-defined Fermi Surface with unprecedented negligible anisotropy for MR.

Isotropic magnetotransport is challenging from the perspective of electron dynamics as mentioned above. A more feasible approach is to couple magnetism with carrier scattering and/or scattering density in an isotropic manner. This approach requires isotropic magnetization which is usually challenging in crystalline materials because magnetic orders are generally characterized by magnetic easy and hard axes. Fortunately, this task is addressable since the mechanism of magnetic anisotropy has been well-explored. Spin-orbital coupling (SOC) and crystal field interactions play key roles in determining magnetic anisotropy of bulk crystalline materials[22]. Electron orbitals are oriented by the lattice due to electrostatic interaction, so that the coupling between electron orbits and spins gives rise to easy-axis or easy-plane anisotropy. Strong SOC and field angle-dependent magnetization can be utilized to create anisotropic MR[20]. Here, minimizing SOC and the crystal field effect may lead to isotropic magnetism. Since this work aims to develop a universal approach to create isotropic magnetotransport, we look into the commonly seen magnetism arising from *d*- or *f*-electrons. *d*-electrons are generally more sensitive to crystal fields. Even though orbital angular momentum for *d* orbitals is usually quenched or partially quenched, SOC still creates residual orbital moment and strong magnetic anisotropy[22,23]. For *f* electrons that are well-screened from the crystal field, however, they are characterized by even stronger SOC and hence significant magnetic anisotropy is expected, which has been known to be a key enabling factor of rare earth-based permanent magnet[22]. An exception applies to the vanishing orbital angular momentum $L$ for half-filled *f*-orbitals, where SOC is effectively zero and the magnetic anisotropy is mainly attributed to shape anisotropy[22]. Therefore, starting from the common +3 valence for rare earth ions, $Gd^{3+}$ with a half-filled $4f^7$ configuration offers the opportunity to realize isotropic magnetism.

Indeed, small magnetic anisotropic energy (MAE) has been found in some Gd compounds such as ferromagnetic GdN (0.45 μeV per Gd) and $GeFe_2$ (9 μeV per Gd)[24], but strong isotropic MR have yet been discovered. The next challenge is to establish an approach to couple transport with magnetism to realize isotropic and strong MR. In magnetic materials, in addition to CMR and GMR in a few materials that do not demonstrate isotropic MR as mentioned above, weak MR due to the field-modulation of spin scattering for charge carriers is more generally seen[25,26]. To realize strong and isotropic MR, field-induced metal-to-insulator (or *vice versa*) transition originating from electronic band reconstruction provides an effective approach. This can be realized via exchange splitting of electronic bands that results in gap closing or opening. Exchange interaction is sensitive to the net moment but not the direction of the moments. Therefore, as illustrated in Fig. 1, starting from an antiferromagnetic (AFM) ground state with zero net moment, finite magnetization induced by magnetic field turns on exchange splitting and modulates electronic bands. For isotropic magnetization, this band structure modulation would be isotropic with respect to the direction of the applied magnetic field. Because *f*-electrons are usually localized, here we pursue exchange interaction between *f* and orbitals near the Fermi energy ($E_F$) from the same $Gd^{3+}$ ions to maximize exchange splitting[27]. Since the exchange interaction splits spin up and down bands, it is easier to start from a gapped system and close the band gap by splitting bands near the $E_F$, as opposed to opening a gap for a metallic system which may involve band structure complications. With this approach, field driven insulator-to-metal transition is expected.

To realize the above scenario, the band gap of the candidate material should be within the band shifts achievable by reasonable exchange splitting, while sufficiently large to minimize the effect of inevitable defects, self-doping, and thermally excited carriers. Therefore, the candidate material should be an AFM semiconductor possessing a moderate band gap, *f*-electron magnetism and *d*-bands near $E_F$. GdPS is a system meeting all the above needs with $Gd^{3+}$ to ensure isotropic magnetism. GdPS was discovered decades ago[28] but only the electronic structure has been explored theoretically[29]. This material adopts a layered crystal structure consisting of flat phosphorus planes sandwiched by Gd-S layers (Fig. 2a). It is structurally similar to the tetragonal ZrSiS[29] with Si square-net planes[30,31], with the difference that GdPS undergoes an orthorhombic structure distortion where the P plane fragments into an arm-chair chain-like structure, as shown in Fig. 2b. Earlier theoretical studies[29] have shown that such a structural distortion plays a vital role in modulating the electronic phase. As shown in Fig. 2c, in tetragonal structure (such as ZrSiS), the $C_{2v}$ symmetry leads to a Dirac nodal-line state that is characterized by linearly dispersed bands crossing near the $E_F$[29,30,31]. In GdPS (Fig. 2d), however, the arm-chair chain configuration of P atoms loses one mirror plane as compared to the square lattice, which strongly gaps the Dirac crossing[29] (see Supplementary Fig. S1 for more details) and eventually leads GdPS to be a semiconductor.

DFT calculations (see Supplementary Information) reveal that GdPS adopts an AFM ground state with a checkerboard (Néel)-type magnetic structure. Though the predicted magnetic structure is difficult to verify by neutron scattering due to the strong neutron absorption of Gd, the AFM ground state is consistent with our magnetization measurements (see Supplementary information). The AFM transition temperature $T_N$ displays a slight sample variation between 7 and 7.5 K, which should be ascribed to the light P vacancies (see Methods). At the AFM ground state, DFT calculations estimate a moderate bandgap of ~0.5 eV (Fig. 2e), which is promising for realizing the desired insulator-to-metal transition as will be shown below.

The vanishing SOC of $Gd^{3+}$, as well as the very weak ligand SOC from the light elements P and S, leads to the needed isotropic magnetism. As shown in Fig. 3a, despite the layered crystal structure, magnetizations measured under in-plane ($H//ab$) and out-of-plane ($H//c$) magnetic fields are almost identical for the temperature below (1.5 K) and above (10 K) $T_N$. The isotropic magnetization persists to higher magnetic fields and saturates above ~15 T for both field orientations, implying a transition from an AFM to a polarized ferromagnetic (FM) state. The saturation moment of ~7.2 $\mu_B$/Gd is consistent with the $^8S^{7/2}$ multiplet for $Gd^{3+}$. The moment polarization persists even above $T_N$, with a nearly unchanged saturation moment at 10 K, as shown in the lower panel of Fig. 3a. Because sample rotation is difficult during high field magnetization measurements, tunnel diode oscillator (TDO) was adopted to examine the moment polarization with field orientations (Fig. 3b). As shown in Fig. 3c, $H_p$ at 0.5 K displays very weak angular dependence. The observed isotropic magnetism in GdPS is consistent with the calculated small MAE of 30 µeV per Gd for the AFM ground state (see Methods). The MAE can be further reduced to only 2 µeV per Gd in the polarized FM state. The combined experimental and theoretical results indicate that the Gd moments rotates freely with the external magnetic field, which is important for the isotropic MR as shown below.

Given the large $Gd^{3+}$ moment, the field-driven spin-polarization leads to strong exchange splitting that modifies the electronic structure. As shown in Fig. 2e, in the AFM ground state, the conduction band edge of the semiconducting GdPS consist of Gd *d*-bands around Γ and P *p*-bands around Y and T points of the Brillouin zone. In the polarized FM state, the strong net moment from Gd sublattice leads to exchange interactions that split both bands, where the Gd *d*-band splitting at the conduction band edge is as large as 0.5 eV (Figs. 2f and 2g, and supplementary Fig. S2) owing to the same-atom *d-f* exchange, as compared to a much smaller band splitting of 0.1 eV for the P *p*-bands. The strong Gd *d*-band splitting at Γ lowers the energy of the spin-polarized band, replacing the P *p*-band edge at Y as the new conduction band minimum that approaches the $E_F$. In real materials, when P vacancies increases $E_F$, a Fermi surface from Gd *d*-band centered at Γ can emerge, leading to an insulator-to-metal transition which has indeed been observed in our magnetotransport experiments described below. Such a process is only determined by the strength of the exchange splitting, thus it is isotropic with respect to magnetic field due to isotropic magnetization of GdPS.

As shown in Fig. 4a, the in-plane resistivity ($\rho_{xx}$) measured with current following parallel to the phosphorus plane of GdPS ($I // ab$) increases with cooling, a typical non-metallic behavior for the gapped electronic structure in the AFM ground state. Applying a perpendicular magnetic field ($H \perp ab$) changes resistivity drastically, and eventually leads to a metallic-like temperature dependence with decreased resistivity upon cooling. The metal state is further verified by quantum oscillations under high magnetic fields as will be discussed later. Such an insulator-to-metal transition is accompanied by gigantic negative MR. At $T = 2$ K, resistivity drops drastically up on increasing field as shown in Fig. 4b. The corresponding MR, quantified by the field-induced resistivity change normalized to its zero-field value, reaches 97.3%. When normalized to the 9T resistivity value following previous reports[15,17,32], the MR is as high as 3,659%. The gigantic negative MR is reproducible in all samples we measured, reaching up to 99% (Supplementary Fig. S3 and Table 1). The strong negative MR occurs in a wide temperature and field region. As shown in Fig. 4b, at 2K, 50% reduction of resistivity is easily achievable under a magnetic field of 3 T, which is much lower than the spin polarization field of 15 T. Also, heating up to 100 K which is well above the magnetic ordering temperature, MR remains as high as nearly 50% at 9 T. This further confirms the exchange splitting scenario that is driven by net magnetization without a need for magnetic ordering or full moment polarization, making GdPS useful for practical applications where operations at high temperature and low field are preferable. As expected, MR in GdPS is isotropic. Figure 4c depicts the angular dependent MR at $T = 5$ K measured by rotating magnetic field both perpendicular to ($\theta$-dependence) and within ($\varphi$-dependence) the phosphorus plane, which reveals that MR is essentially independent of the field orientation. This isotropic MR is reproducible and persists at high temperatures beyond $T_N$, as shown in the Supplementary Fig. S4.

To better quantify the isotropy of MR, we define the isotropic index by the difference of MR maximum and minimum in the angular dependent measurements, normalized to the average MR (see Methods). Such definition takes the strength of MR into account so that it characterizes both isotropy and amplitude of MR. The calculated isotropic index is as small as 0.8% and 0.2% at 9T for out-of-plane ($\theta$-) and in-plane ($\varphi$-) rotation, respectively.

Similar isotropic and gigantic MR is also present in out-of-plane electronic transport, where the electrical current following perpendicularly to the phosphorus plane, as shown in Figs. 4d-e. Compared to in-plane transport, the zero-field out-of-plane resistivity $\rho_{zz}$ at low temperatures is much larger than $\rho_{xx}$ by a factor of 35 (Fig. 4d), consistent with the layered crystals structure of GdPS. However, $\rho_{zz}$ is much more sensitive to magnetic field and drops strongly under a small magnetic field. As shown in Fig. 4e, at 2 K, negative MR up to 90% for $\rho_{zz}$ can be induced by a small field of 1 T. Further increasing field to 9 T boosts MR to 99.99%, which converts to 740,167% when normalizing to the 9 T resistivity value. Similar to $\rho_{xx}$, strong negative MR in $\rho_{zz}$ persists to high temperatures with a much greater value of 93.8% at 100 K and 9 T (Fig. 4e), and displays isotropic field orientation dependence with isotropic index < 0.01% for both out-of-plane and in-plane rotations (Fig. 4f). The isotropic MR in $\rho_{zz}$ also persists at high temperatures, as shown in the Supplementary Fig. S5.

The isotropic and gigantic MR in both in-plane and out-of-plane magnetotransport drastically differs from other strong but anisotropic MR compounds such as EuMnSb$_2$[15,16], EuTe$_2$[17,18], Mn$_3$Si$_2$Te$_6$[19,20], and EuCd$_2$P$_2$[21]. The only known example of large, isotropic MR material is a disordered topological insulator with delicate composition tuning to allow a formation of a percolation of the current paths under field, and hence the MR isotropy is sample dependent[33]. Therefore, GdPS represents the first example of isotropic gigantic MR in pristine crystals which is driven by a well-defined insulator-to-metal transition from exchange splitting. The isotropic magnetism of GdPS ensures isotropic MR at various temperatures (above or below $T_N$) and magnetic fields, which is distinct from other known materials showing strong negative MR as summarized in Supplementary Fig. S6. Such a unique property again highlights the potential of GdPS for applications.

Quantitatively comparing in-plane and out-of-plane transport at low temperatures, one can find that $\rho_{zz}$ is much larger at 0 T but becomes comparable with $\rho_{xx}$ at 9 T. This suggests that electronic transport becomes more isotropic under a magnetic field, which can be understood by the evolution of electronic structure during the insulator-to-metal transition. Electronic structure at zero magnetic field can be probed by angle-resolved photoemission spectroscopy (ARPES). As shown in Figs. 5a-c, ARPES observations are qualitatively consistent with the calculated band dispersions in Fig. 2e. At the Y point, a small electron pocket is seen, which can be attributed to P vacancies that electron-does GdPS. Despite its presence, this electron pocket is not large enough to support full metallicity, as the non-metallic transport up to room temperature at zero field (Fig. 4a) implies our GdPS is a lightly doped semiconductor.

The electronic structure under magnetic fields, though unattainable by ARPES, can be resolved by quantum oscillations. Both Shubnikov–de Haas (SdH) oscillations in magnetotransport (Fig. 5e) and de Haas-van Alphen (dHvA) oscillations in TDO measurements (Fig. 3b and Supplementary

Fig. S9d-f) have been observed in our high field experiments. One typical example is shown in Fig. 5e, from which SdH effects in MR data can be clearly seen above ~17 T at $T$ = 0.5 K. Importantly, SdH oscillation is observed for any magnetic field orientations ranging from out-of-plane ($\theta$ = 0°) to nearly in-plane ($\theta$ = 82.6°) directions, which indicates a well-defined 3D Fermi surface and rule out the possibility of quantum oscillation from surface states in insulators[34,35] The oscillation frequency does not vary strongly with field orientations, growing by only 16% from 500 T for $\theta$ = 0° to 580 T for $\theta$ = 82.6°, as shown in Fig. 5f. Since the oscillation frequency is proportional to the extremal cross-sectional area perpendicular to the magnetic field orientation, such angular dependence implies a nearly spherical Fermi surface, which is consistent with the 3D Fermi surface at Γ from the DFT band structure. As discussed above (Fig. 2f), in the polarized FM state, the strong $d$-$f$ exchange splitting may lead to a Fermi surface at Γ from the spin polarized Gd $d$-band. Indeed, assuming a shifted $E_F$ by 0.245 eV attributed to self-doping from the P vacancies, the calculated angular dependence of oscillation frequency matches well with the experimental observations, as shown in Fig. 5f. Aside from the Gd $d$-band, other possible bands such as the P $p$-band are not resolvable by quantum oscillations.

The nearly spherical Fermi surface at Γ emerging with increasing magnetic fields is further supported by the surprising quantum oscillation anomalies. Figure 5g depicts the contour plot showing the temperature and field dependencies of the SdH effect under an in-plane magnetic field. Interestingly, the locations of the oscillation maxima and minima shift systematically with varying temperature, as illustrated by the bent stripes in Fig. 5g. Typical data taken at 2 K and 12 K are shown in the inset, where misaligned oscillations are clearly seen. Such behavior is observed in multiple samples, occurring in both SdH and dHvA effects (Supplementary Fig. S9), which is caused by the temperature dependent oscillation frequency that drops from 560 T at 1.5K to 450 T at 18K, as shown in Fig. 5h. This appears inconsistent with the standard quantum oscillation theory based on Landau quantization of electronic systems with unchanged charge densities, but agrees well with the scenario of exchange splitting-driven insulator-to-metal transition: The emergence of Fermi surface is due to the exchange-split Gd $d$-bands that crosses the $E_\mathrm{F}$, hence the size of the Fermi surface is determined by the strength of the sample magnetization which reduces upon warming. Similarly, quantum oscillation frequency also evolves with magnetic field in the same manner as magnetization. As shown Fig. 5i, the oscillation frequency at 1.5 K grows with field with a saturation behavior. At elevated temperatures (6 and 10 K), the frequency is lower but the field dependence becomes even stronger. All of these match well with the magnetization of GdPS (Fig. 3a). With the electronic structures revealed by ARPES and quantum oscillation, the much stronger MR in out-of-plane transport can now be understood. At zero magnetic field, transport is governed by P layers as revealed by ARPES observations, so that $\rho_{zz}$ is much greater than $\rho_{xx}$ in layered GdPS. Applying magnetic field induce a 3D spherical-like Fermi surface from Gd $d$-bands, so that $\rho_{zz}$ drops strongly and eventually the out-of-plane transport becomes comparable with the in-plane transport.

The layered crystal structure of GdPS allows exfoliation down to thin flakes (Supplementary Fig. S10). Thus, this material offers an excellent platform for device integration, especially for omnidirectional sensing. The AFM ground state with vanishing stray fields further enables ultrafast and high-efficient device response[36,37]. In addition, in the field-induced metallic state, the Fermi surface formed from the exchange-split Gd $d$-band is spin polarized. Therefore, controlling $E_\mathrm{F}$ by chemical or electrostatic doping to remove possible contribution from the phosphorus bands may lead to a fully spin polarized Fermi surface, which opens vast opportunities for spintronic applications. Our finding highlights the effectiveness of utilizing exchange splitting-driven insulator-to-metal transition to realize isotropic and gigantic modulation of transport by magnetic field, which is enabled by suppressed crystal field effect in *f*-electron systems and near-zero SOC in half-filled *f*-orbitals to create isotropic magnetism, as well as an AFM and semiconducting ground state. Such a strategy can be generalized to discover more candidate materials by computational high-throughput screening followed by experimental optimization of material parameters by doping or strain, to ensure band edge states by *d*-orbitals experiencing strong *d-f* exchange, as well as appropriate spin polarization fields and magnetizations at elevated temperatures to extend the phenomena towards high temperatures and low fields for technology applications.

**Methods**

**Sample preparation**

Single crystals of GdPS were grown by a direct chemical vapor transport with a stoichiometric ratio of Gd, P, and S as source materials and $TeCl_4$ as the transport agent. The growths were performed in a dual heating zone furnace with a temperature gradient from 1075 to 975 °C for 3 weeks. The produced crystals typically have either needle-like or plate-like structures (see Supplementary information), both of which were confirmed to be the GdPS phase by the composition analysis using energy-dispersive x-ray spectroscopy and the structure analysis using x-ray diffraction.

The structure of the GdPS single crystal was determined by single crystal XRD. The structure parameters obtained from single crystal XRD refinement were given in Supplementary information. A careful structure investigation found that the atomic displacement parameter of P atoms (0.0116(10) $Å^2$) is greater than other atoms including Gd1 (0.0065(5) $Å^2$), Gd2 (0.0070(5) $Å^2$), S1 (0.0060(12) $Å^2$), and S2 (0.0050(12) $Å^2$) under the same refinement conditions, which indicated the possible presence of vacancy at P atomic sites. The refinement of atomic occupancy of P atoms revealed the occupancy of 0.95, which indicates the presence of P vacancy that is consistent with the electron doping scenario revealed by ARPES.

**Angle-resolved photoemission spectroscopy**

ARPES measurements were conducted out at Beamline 10.0.1 of the Advanced Light Source (ALS) in Berkeley, California, USA. The energy resolution was better than 30 meV and the temperature for measurement was 90 K. The incident photon energy was 55 eV. Samples were cleaved in situ and measured under vacuum better than 5 × 10-11 Torr.

**Band structure calculations**

The band structure was calculated using density-functional theory (DFT) as implemented in VASP[38] using the Perdew-Burke-Ernzerhof (PBE) exchange-correlation functional[39] and projector augmented wave (PAW) pseudo-potentials. The cut-off for the kinetic energy of plane was to be 520 eV and a 4x4x1 k-point mesh was used in the Brillouin zone. We used the Hubbard U parameter for the Gd-4$f$ electron ($U_{eff}^{Gd,5f}$ = 7 eV). The total energy was converged at least $10^{-6}$ eV ($10^{-7}$ eV for MAE) during electron configuration relaxation. Spin-orbit coupling (SOC) was included in all calculations unless otherwise stated. To model the paramagnetic state for GdPS (Supplementary Fig. S1), the calculation was performed by using a non-magnetic state.

**Magnetic anisotropy energy calculations**

The calculation of the variation of MEA with polar angle $\theta$ and azimuthal angle $\varphi$ provides stable structures as references for MAE calculations. As shown in Supplementary information, we observed that in the case of AFM GdPS, the configuration with the magnetic moment oriented at $\theta$ = 45° and $\varphi$ = 45° is slightly more stable than other AFM configurations. Similarly, in the case of FM GdPS, the configuration with the magnetic moment oriented at $\theta$ = 0° and $\varphi$ = 0° is slightly more stable than other FM configurations. We consider these stable structures as references for MAE calculations.

**Extremal Fermi surface cross-section area and associated quantum oscillation frequency**

To calculate the Fermi surface with the interpolation of the energy bands, the maximally localized Wannier functions (MLWF) for the ferromagnetic state of GdPS were obtained using the Wannier90 package[40]. Moreover, the dHvA frequencies were calculated by the SKEAF code[41] with the Fermi surface information.

**Measurements**

The magnetization and magnetotransport up to 9 T were measured using a physical property measurement system (PPMS, Quantum Design). The standard 4-probe measurement was used for magnetotransport. The high field magnetization measured with VSM (up to 35 T), magnetotransport measurements (up to 31 T and 45 T), and TDO measurements (up to 45 T) were performed at the National High Magnetic Field Laboratory (NHMFL) at Tallahassee.

**Anisotropic index**

Anisotropic index characterizes the degree of anisotropy. It is defined as the difference between the maximal and minimal MR at different magnetic field orientations $[MR(\theta)_{max} - MR(\theta)_{min}]$, normalized by the average MR $[MR(\theta)_{avg}]$ calculated by averaging the maximal and minimal MR:

$$\text{Anisotropic index} \equiv \frac{MR(\theta)_{max} - MR(\theta)_{min}}{MR(\theta)_{avg}} = \frac{MR(\theta)_{max} - MR(\theta)_{min}}{[MR(\theta)_{max} + MR(\theta)_{min}]/2}$$

Therefore, anisotropic index takes a value between 0 and 2. A value of 0 means perfect isotropic when $MR(\theta)_{max} = MR(\theta)_{min}$, and a value of 2 refers to completely anisotropic when $MR(\theta)_{min} = 0$.

MR changes with temperature and magnetic field, which allows for extracting the anisotropic index as a function of the amplitude of MR, as shown in Supplementary Fig. S6. Other materials with strong negative MR, including EuMnSb$_2$[15], EuTe$_2$[18], Mn$_3$Si$_2$Te$_6$[19], EuCd$_2$P$_2$[21], disordered topological insulator TlBi$_{0.15}$Sb$_{0.85}$Te$_2$[33], and CeCuAs$_2$[42] are provided in Supplementary Fig. S6 for comparison.

**Extracting Temperature- and Field-dependent quantum oscillation frequency**

Because the Fermi surface of the polarized FM state is formed from the exchange-split Gd *d*-bands, it evolves with magnetization and thus temperature- and field-dependent. Therefore, the corresponding quantum oscillation frequency, which is proportional to the extremal Fermi Surface cross-section area perpendicular to magnetic field direction, also changes with temperature and magnetic field. For field-dependence presented in Fig. 5i, the oscillation frequency cannot be extracted from the conventional Fourier transform because the oscillation data is in the field-domain. Therefore, to determine oscillation frequency at various magnetic fields, we evaluate the distance between neighboring oscillation peaks, which is the oscillation period. The inverse of the period yields frequency for that field range, as shown in Extended Data Fig. 8c.

For temperature dependence, the oscillation frequency is obtained from the Fourier transform of the oscillation pattern, as shown in Extended Data Fig. 8b, which is the convention method for quantum oscillation analysis. For temperature-dependent Fermi surface in GdPS, this method bears small errors. That is because Fourier transform is applied to a certain field range containing multiple oscillation periods, but the frequency changes with field in GdPS. Nevertheless, given that the oscillation amplitude damps with reducing magnetic field, the peak position in Fourier spectrum is mainly determined by the oscillation in the high field range. Indeed, frequency extracted from Fourier transform in Fig. 5h agrees with the high field (29 – 31T) frequency obtained by the distance between neighboring oscillation peaks in Fig. 5i.

**Data availability**

All data that support the findings of this study are available from the corresponding authors on request. Source data are provided with this paper.

**Acknowledgements**

J.H acknowledges the support by the U.S. Department of Energy, Office of Science, Basic Energy Sciences program under Grant No. DE-SC0022006 for crystal growth, transport and magnetic property measurements. J. H. thanks Z. Q. Mao at PennState and X.G. Zhang at UF for informative discussions. B.N. and Y.W. acknowledge startup funds from the University of North Texas, and computational resources from the Texas Advanced Computing Center. Part of the modeling was supported by computational resources from a user project at the Center for Nanophase Materials Sciences (CNMS), a US Department of Energy, Office of Science User Facility at Oak Ridge National Laboratory, and also partially by user project R0076 at the Pennsylvania State University Two-Dimensional Crystal Consortium – Materials Innovation Platform (2DCC-MIP) under NSF cooperative agreement DMR-2039351. A portion of this work (high field transport, TDO, and magnetization) was performed at the National High Magnetic Field Laboratory, which is supported by National Science Foundation Cooperative Agreement No. DMR-2128556 and DMR-2128556 and the State of Florida. Work at Nanyang Technological University was supported by the National Research Foundation, Singapore, under its Fellowship Award (NRF-NRFF13-2021-0010), the Agency for Science, Technology and Research (A*STAR) under its Manufacturing, Trade and Connectivity (MTC) Individual Research Grant (IRG) (Grant No.: M23M6c0100), Singapore Ministry of Education (MOE) AcRF Tier 2 grant (MOE-T2EP50222-0014) and the Nanyang Assistant Professorship grant (NTU-SUG). Work at Princeton University was supported by the Gordon and Betty Moore Foundation (GBMF4547 and GBMF9461; M.Z.H.). The ARPES work was supported by the US DOE under the Basic Energy Sciences program (grant number DOE/BES DE-FG-02-05ER46200; M.Z.H.). This research used resources of the Advanced Light Source, which is a DOE Office of Science User Facility under contract no. DE-AC02-05CH11231. The authors want to thank S.-K. Mo at Beamline 10.0.1 of the ALS for support in getting the ARPES data. M.Z.H. acknowledges support from Lawrence Berkeley National Laboratory and the Miller Institute of Basic Research in Science at the University of California, Berkeley in the form of a Visiting Miller Professorship. M.Z.H. also acknowledges support from the U.S. Department of Energy, Office of Science, National Quantum Information Science Research Centers, Quantum Science Center. Work at Wichita (single crystal XRD and structure refinement) is supported by NSF under award OSI-2328822. Work at NIMS is supported by JSPS KAKENHI Grant Number JP21K14656, the Kurata Grants from The Hitachi Global Foundation and from The Iketani Science & Technology Foundation.

**Author contributions**

G.A, B.N, C.-H.H, and X.P.Y contribute to this work equally. G.A, K.P., S.K.C, and R.B grew the crystals. G.A, R.B., S.R, H.C performed electrical transport and low field magnetotransport experiments. G.A, M.R.U.N, S.K.C, D.G., and E.S.C conducted high field magnetotransport, TDO, and magnetization measurements. J.W., B.D, and Z.H performed structure analysis. B.N, Y.W. C.-H.H, and G.C performed computations. X.P.Y. and M.Z.H. performed ARPES investigations. J.H. and Y.W. conceived the project.

# Figures

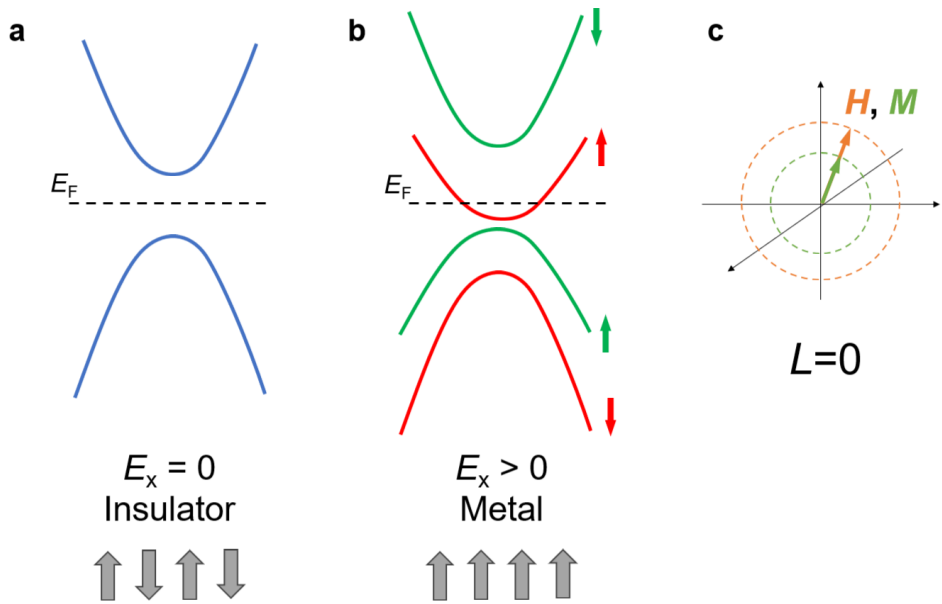

**Figure 1 | Strategy for realization isotropic magnetoresistance. a,** Schematic band structure for the ground state (zero magnetic field) of an AFM semiconductor insulator, showing a gap opening at the Fermi energy $E_F$. Exchange energy $E_x$ is zero due to the vanished net magnetic moment in AFM state, as illustrated in the bottom. **b**, Applying magnetic field polarizes the moments to form a FM state, creating strong net moment and turn on exchange interaction that split the band. Sufficiently strong exchange splitting drives a band to cross the $E_F$, resulting in a metallic state. **c**, Isotropic magnetization. The magnetization (***M***) of the sample is induced and parallel to the external magnetic field (***H***), rotating in-phase with the field rotation.

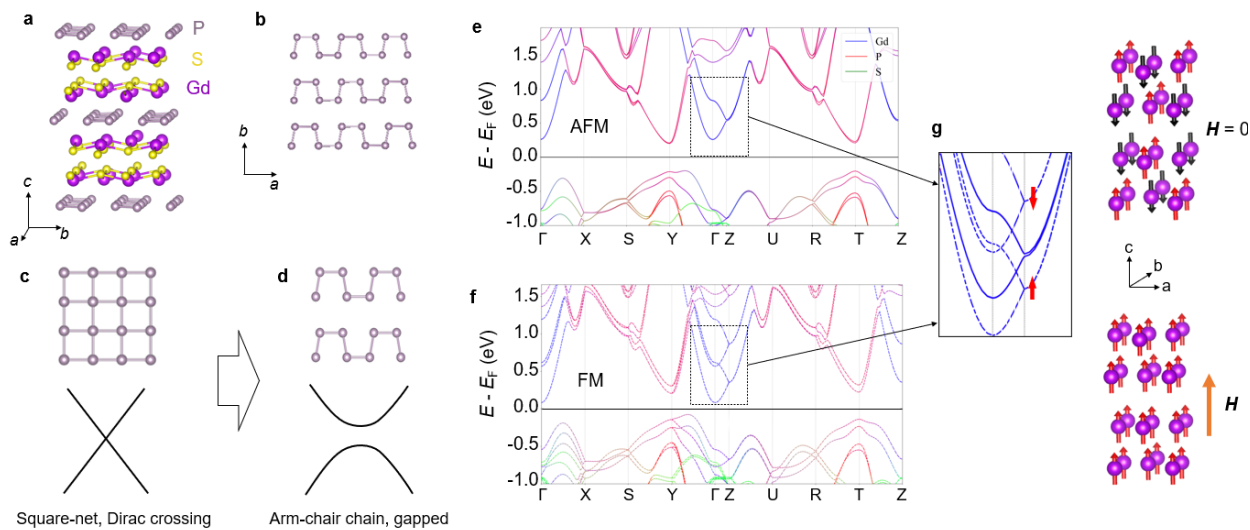

**Figure 2 | Electronic structure of GdPS. a**, the orthorhombic crystal structure of GdPS. **b**, arm-chair chain structure of the P layers. **c**, the tetragonal structure with square-net layer leads to Dirac crossings. **d,** the orthorhombic distorted arm-chair chain-type P layer opens a gap at the Dirac point (see Extended Data Fig. 1 for detailed band structures). **e,** Electronic band structure of GdPS in the AFM ground state. The right panel shows the checkerboard (Néel)-type AFM magnetic structure for the ground state determined by calculations (see Supplementary information). While the magnetic structure is determined, the Gd moments' orientations are schematic, which is difficult to determine owing to the very weak MAE. **f,** Electronic band structure of GdPS in the polarized FM state. Magnetic structure for the FM state shows the polarization of Gd moments along the direction of magnetic field. **g,** Exchange splitting between spin up and down bands for the Gd $d$-band near Γ, which can be seen by overlaying the electronic structures of the AFM state (solid line) and the FM state (dashed line). The comparison of the entire electronic band structure is shown in Supplementary Fig. S2.

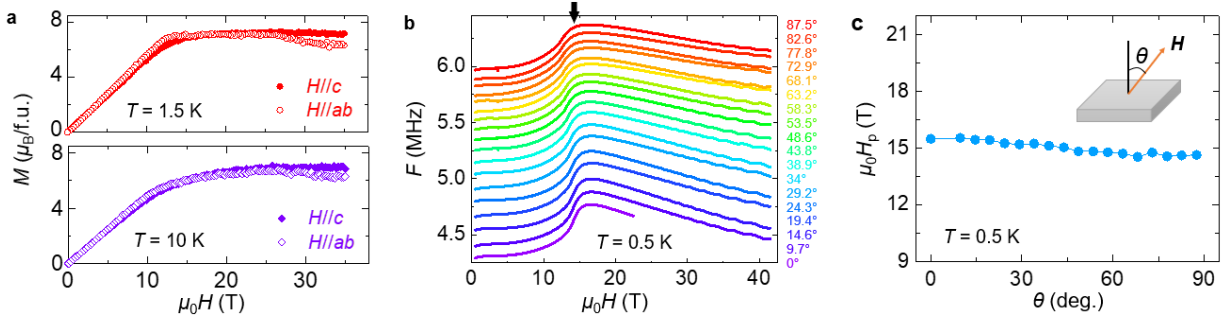

**Figure 3 | Magnetic properties of GdPS. a**, Field-dependent magnetization of GdPS at (top) 1.5 K and (bottom) 10 K, measured with field along the *c*-axis (perpendicular to the P plane) and within the *ab*-plane (parallel to the P plane). Magnetization exhibit saturation behavior at high field, indicating a polarized FM state. **b**, TDO measured at various magnetic field orientations. The transition-like feature near 15 T (indicated by the black arrow) signatures the moment polarization to an FM state. At high field, dHvA oscillation in TDO can be observed. **c**, Critical field for moment polarization, as a function of the magnetic field orientation, extracted from TDO measurement. The polarization field $H_\mathrm{p}$ is essentially field-orientation-independent.

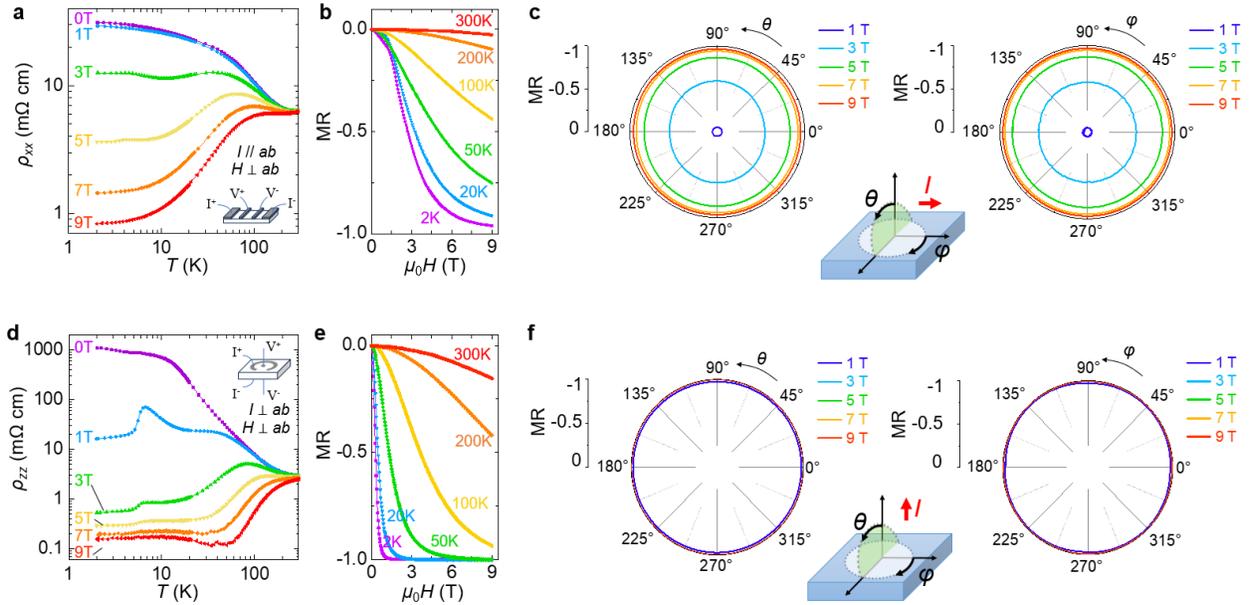

**Figure 4 | Gigantic, isotropic negative MR in GdPS. a**, Temperature dependence for in-plane resistivity $\rho_{xx}$, measured under various magnetic field from 0 to 9 T. Inset: measurement setup. **b**, Magnetic field dependence for $\rho_{xx}$, measured at various temperatures from 2 to 300 K. **c**, Angular dependence for MR at $T = 5$ K for in-plane resistivity $\rho_{xx}$, measured by rotating field perpendicular to ($\theta$-dependence, left) and within ($\varphi$-dependence, right) the phosphorus plane. Inset: schematic of magnetic field rotation. **d**, Temperature dependence for out-of-plane resistivity $\rho_{zz}$, measured under various magnetic field from 0 to 9 T. Inset: measurement setup. **e**, Magnetic field dependence of $\rho_{zz}$, measured at various temperatures from 2 to 300 K. **f**, Angular dependence for MR at $T = 5$ K for out-of-plane resistivity $\rho$, measured by rotating field perpendicular to ($\theta$-dependence, left) and within ($\varphi$-dependence, right) the phosphorus plane. Inset: schematic of magnetic field rotation.

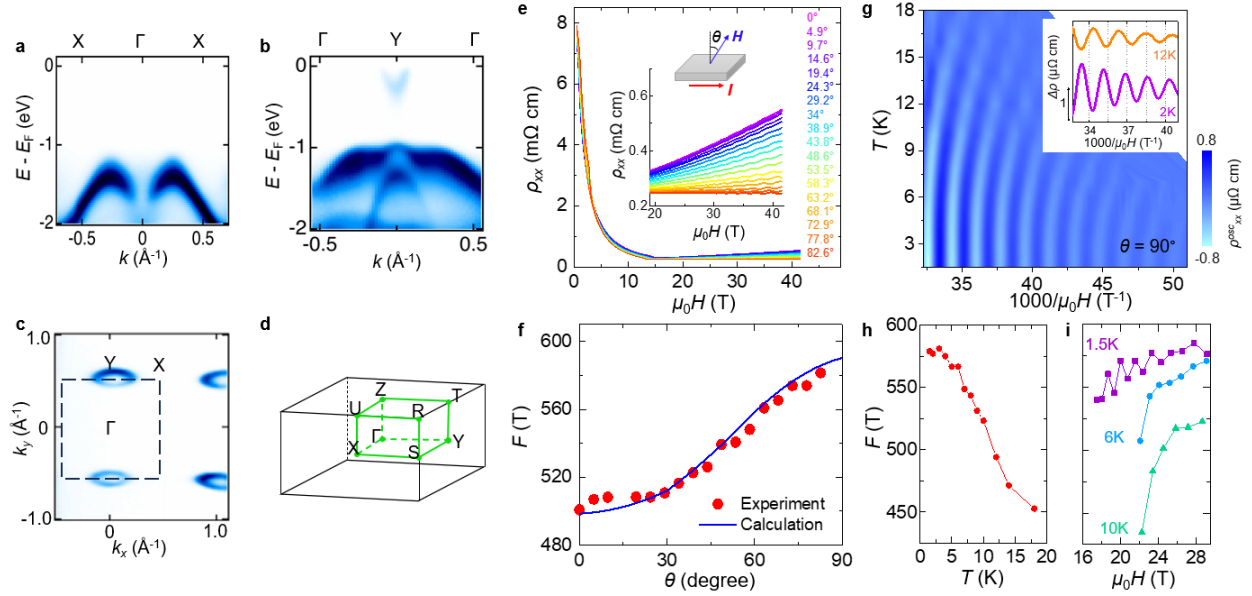

**Figure 5 | Electronic structure of GdPS. a-d**, **Electronic band structure at zero field**. **a-b**, APRES observation of the electronic band dispersion at 90 K along (**a**) Γ-X and (**b**) Γ-Y directions. An electron band slightly crosses the $E_F$ at Y, leading to a small Fermi surface at Y point as shown in **c**. **d**, Brillouin zone for GdPS. The black dotted square in **c** marks the surface Brillouin zone. **e**, Angular dependence for MR for out-of-plane resistivity $\rho$, measured by rotating field perpendicular to ($\theta$-dependence, left) and within ($\varphi$-dependence, right) the phosphorus plane. Inset: schematic of magnetic field rotation. **e-i**, **Electronic band structure of the polarized FM state**. **e**, High field MR at $T = 0.5$ K for GdPS, measured at various magnetic field orientations. Upper inset shows the measurement schematic. Lower inset shows the SdH oscillation at high field. **f**, Field-angular dependence for SdH oscillation frequency obtained from data in **e** (see Supplementary Fig. S7). The observed angular dependence matches well with the calculated Fermi surface formed from the exchange-split Gd $d$-band at Γ, assuming shifted $E_F$ by 0.245 eV (see Methods). **g**, Contour plot showing the dependence of the SdH oscillatory component $\rho_{xx}^{osc}$ with temperature and inverse magnetic field, constructed from SdH effect at various temperatures shown in Supplementary Fig. S8a). Inset: Oscillation pattern at 2 and 12K. **h-i**, (h) Temperature- and (**i**) field-dependence for SdH oscillation frequency of the same sample (see Methods for extracting frequencies). Results in **e-i** are from the identical sample measured using (**e-f**) a 41 T magnet and (**g-i**) a 31 T magnet.

# Supplementary Information for

# Insulator-to-Metal Transition and Isotropic Gigantic Magnetoresistance in Layered Magnetic Semiconductors

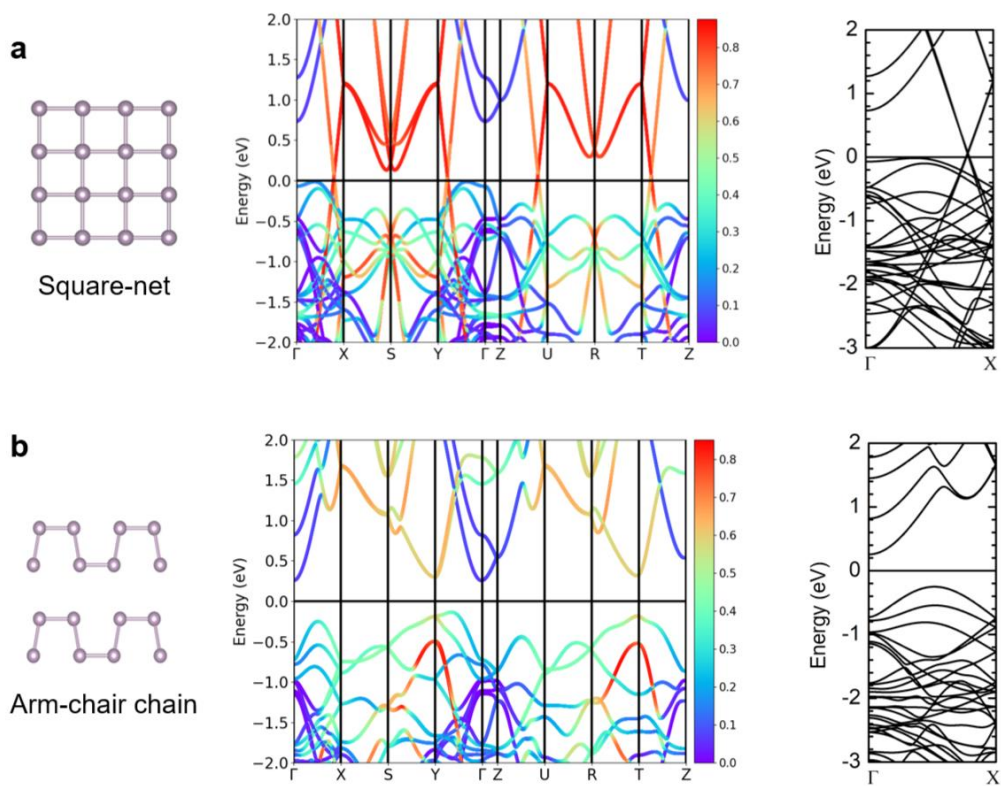

**Fig. S1 | Symmetry breaking gap in GdPS.** Calculated electronic band structure in the non-magnetic state (see Methods) for (**a**) a tetragonal ZrSiS-type structure with square nets of P-layers, and (**b**) an orthorhombic structure with arm-chair chain of P-layers. The middle panels show the electronic band structure with the intensity indicating contribution from P. The right panels display zoom-in of the Γ-X cut to highlight the band opening at the Dirac crossings at the $E_F$. SOC is included in the calculation. Contribution from the Gd-$f$ orbital is not counted for modeling the non-magnetic state.

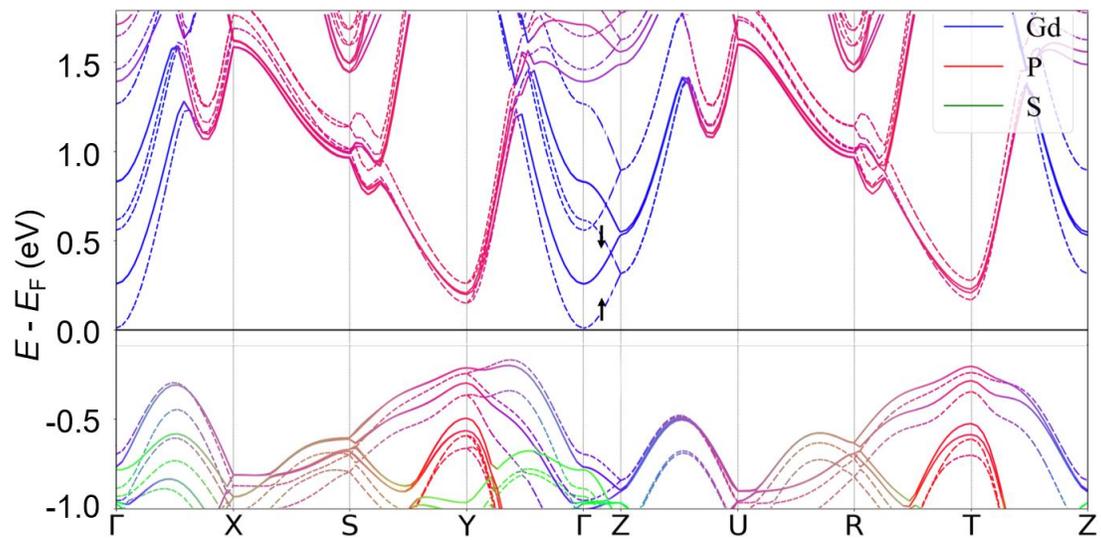

**Fig. S2 | Exchange splitting in GdPS.** Overlay of the electronic structures of the AFM ground state (solid lines) and the polarized FM state (dashed lines). The energies of the two band structures are aligned by the location of the unsplit bands.

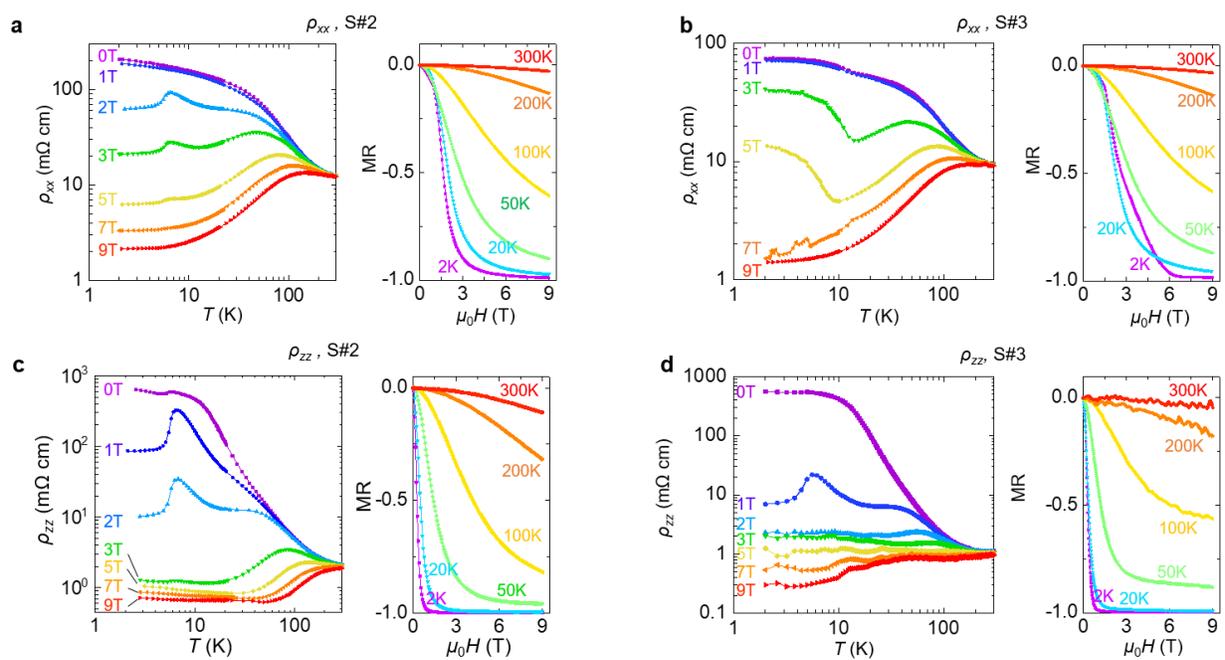

**Fig. S3 | Reproducible gigantic, isotropic negative MR in multiple samples. a-b,** Temperature and magnetic field dependence for in-plane resistivity $\rho_{xx}$ for another two samples. **c-d,** Temperature and magnetic field dependence for out-of-plane resistivity $\rho_{zz}$ for another two samples.

**Table S1 | Negative MR of various samples**

| Samples | Source | Resistivity at 2K, 0T, $\rho(H=0)$ (mΩ cm) | Resistivity at 2K, 9T, $\rho(H=9T)$ (mΩ cm) | MR at 2K, normalized to $\rho(H=0)$ $\frac{\rho(H=0)-\rho(H=9T)}{\rho(H=0)}$ | MR at 2K, normalized to $\rho(H=9T)$ $\frac{\rho(H=0)-\rho(H=9T)}{\rho(H=9T)}$ |
|---|---|---|---|---|---|
| $\rho_{xx}$, S#1 | Fig. 4a | 31.2 | 0.83 | 97.3% | 3,659% |
| $\rho_{xx}$, S#2 | Fig. S3 | 208.2 | 2.1 | 99.0% | 9,814% |
| $\rho_{xx}$, S#3 | Fig. S3 | 74.3 | 1.4 | 98.1% | 5,207% |
| $\rho_{zz}$, S#1 | Fig. 4d | 1,110.4 | 0.15 | 99.99% | 740,167% |
| $\rho_{zz}$, S#2 | Fig. S3 | 640 | 0.72 | 99.9% | 88,788% |
| $\rho_{zz}$, S#3 | Fig. S3 | 560 | 0.3 | 99.9% | 185,667% |

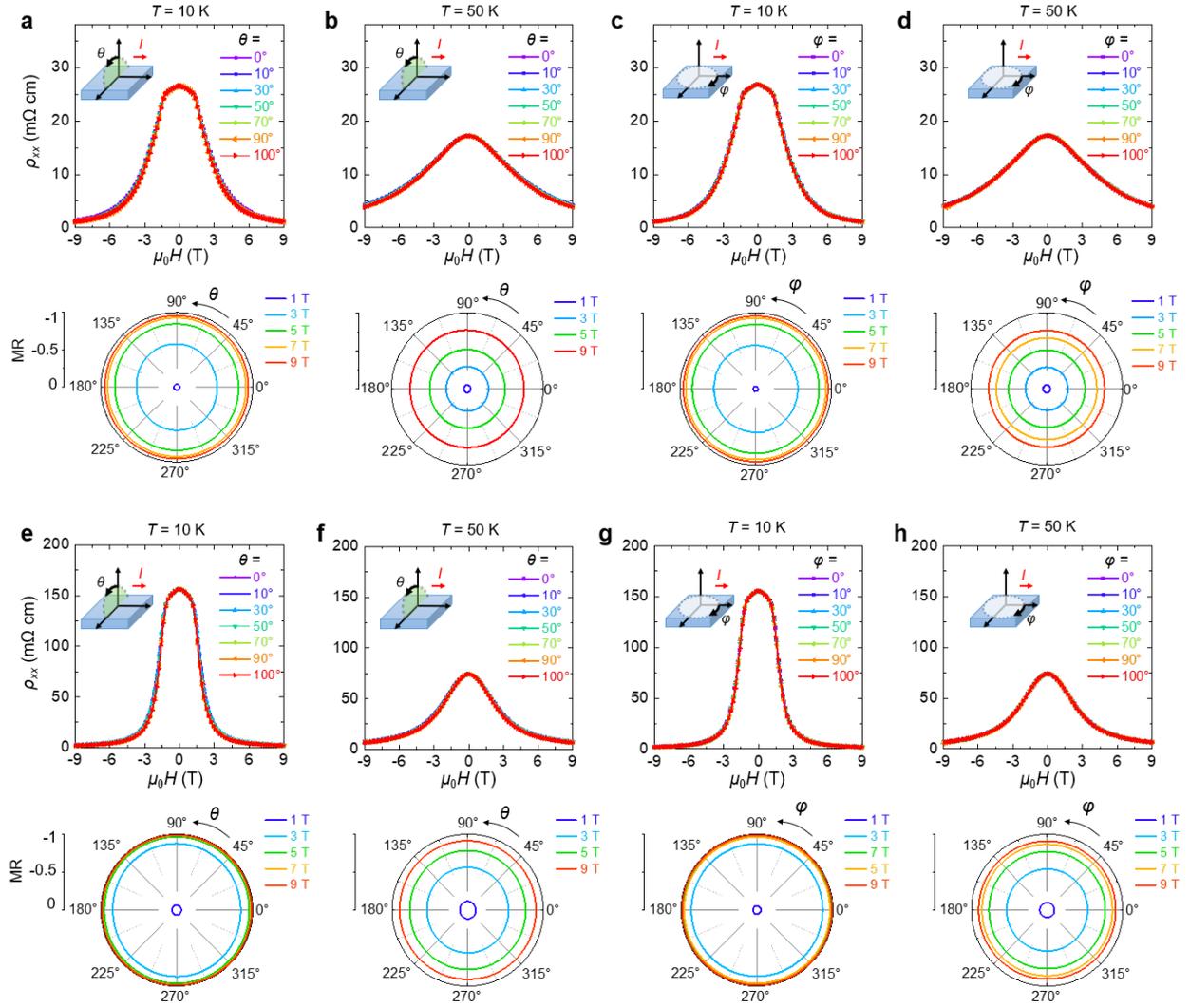

**Fig. S4 | Reproducible gigantic, isotropic negative MR at high temperatures for in-plane transport. a-b**, Top: field dependence for in-plane resistivity $\rho_{xx}$ at (**a**) $T = 10$ K and (**b**) $T = 50$ K, measured by applying magnetic field along different directions perpendicular to the phosphorus plane of GdPS (various $\theta$ angles, see inset). Bottom: corresponding angular dependence for MR at different fields at 10 and 50 K. **c-d**, similar measurements to **a-b** for the identical sample, but with various in-plane field orientations ($\varphi$ angles, see inset). **c-d**, similar measurements to **a-d**, but for a different sample. The two samples display different amplitudes for MR, but both are gigantic and isotropic.

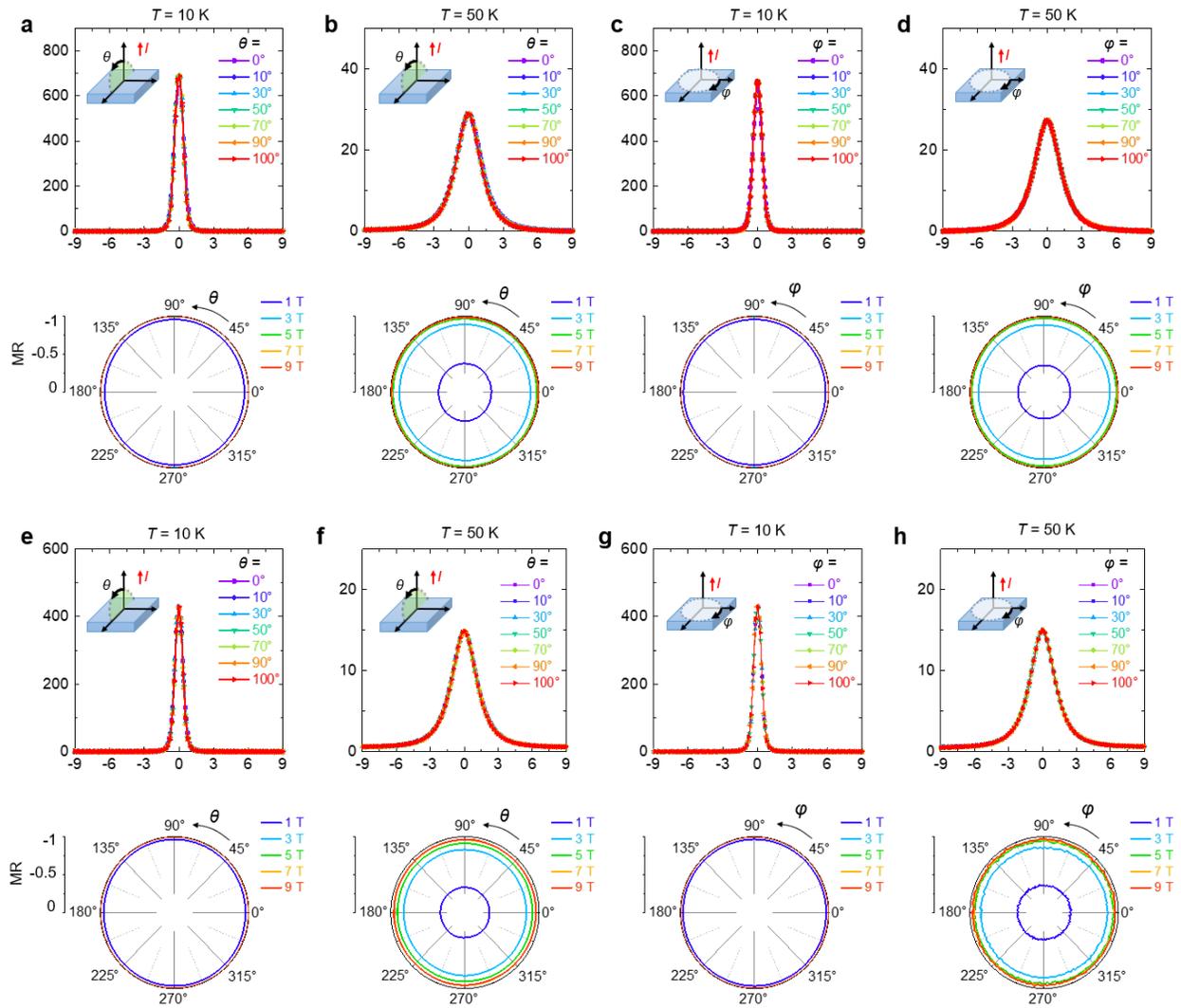

**Fig. S5 | Reproducible gigantic, isotropic negative MR at high temperatures for in-plane transport. a-b**, Top: field dependence for in-plane resistivity $\rho_{zz}$ at (**a**) $T$ = 10 K and (**b**) $T$ = 50 K, measured by applying magnetic field along different directions perpendicular to the phosphorus plane of GdPS (various $\theta$ angles, see inset). Bottom: corresponding angular dependence for MR at different fields at 10 and 50 K. **c-d**, similar measurements to **a-b** for the identical sample, but with various in-plane field orientations ($\varphi$ angles, see inset). **c-d**, similar measurements to **a-d**, but for a different sample. The two samples display different amplitudes for MR, but both are gigantic and isotropic.

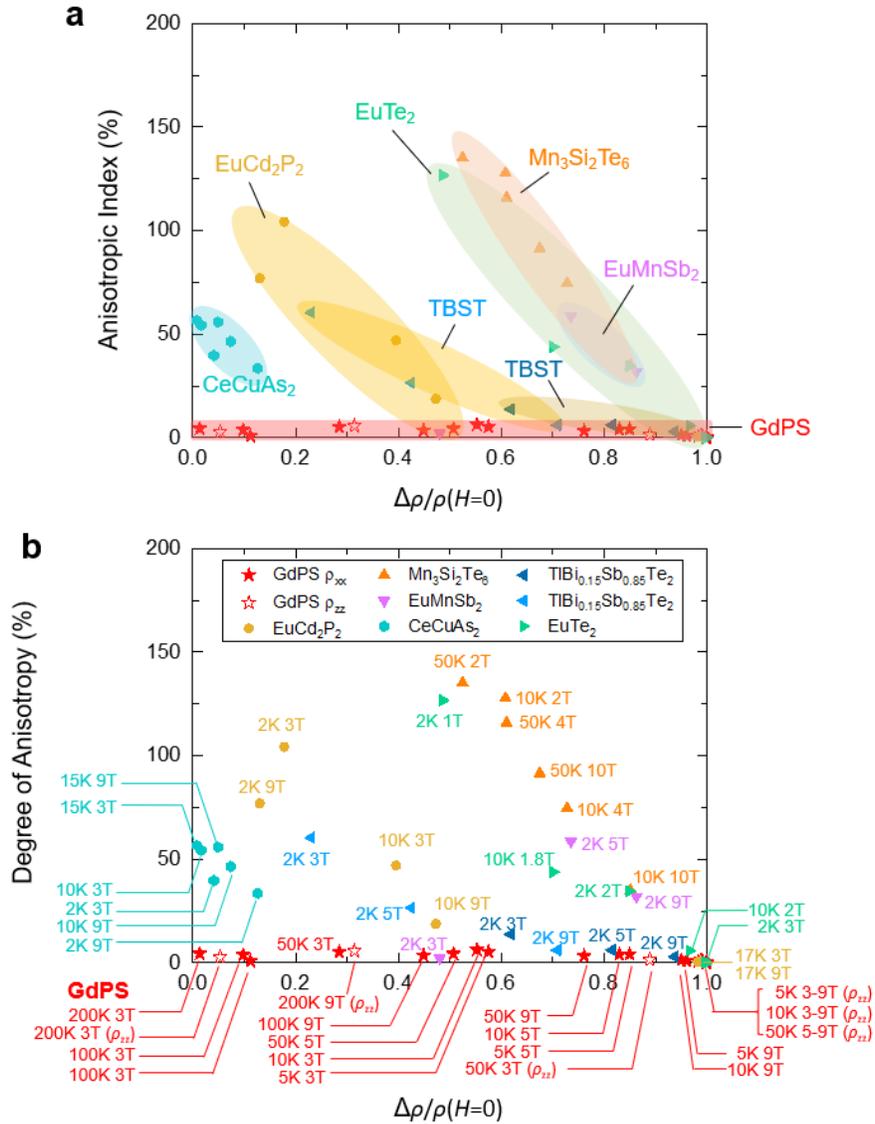

**Fig. S6 | MR anisotropy characterized by the anisotropic index. a,** Anisotropic index as a function of MR (see Methods). Because that MR measures the relative change of a physical quantity and thus it represents the change as compared to the initial value (the zero-field resistivity), here MR is normalized to the zero-field resistivity $\Delta\rho/\rho(H=0)$ and cannot exceed 100%. Other known materials with strong negative MR are provided for comparison. The two TBST represents two $TlBi_{0.15}Sb_{0.85}Te_2$ samples in the same report. **b,** The same plot with labelled temperature and magnetic field for each data point. All other materials show substantial MR anisotropy at some temperatures and magnetic fields. In contrast, GdPS displays isotropic MR in the entire temperature and magnetic field range, and is not affected by AFM transition at 7 K.

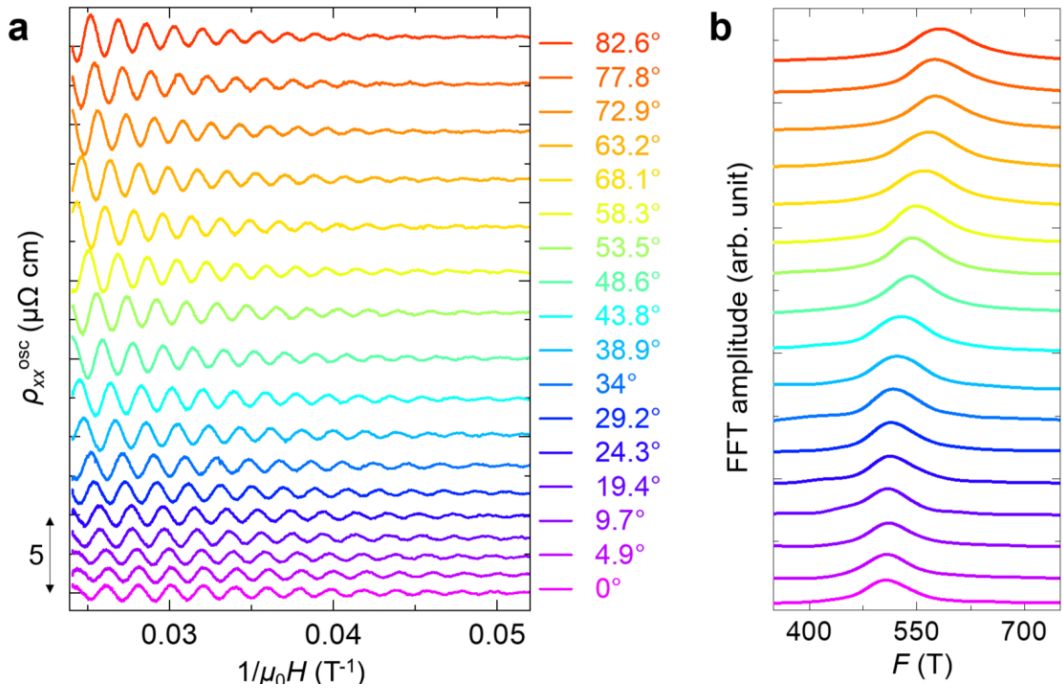

**Fig. S7 | Field-angle dependent quantum oscillation. a,** The oscillatory components of SdH oscillations in $\rho_{xx}$ under various magnetic field orientations presented in Fig. 5e, extracted by removing the non-oscillatory MR background. **b**, Fast Fourier transform (FFT) for the oscillatory components in **a**. Data are shifted for clarity.

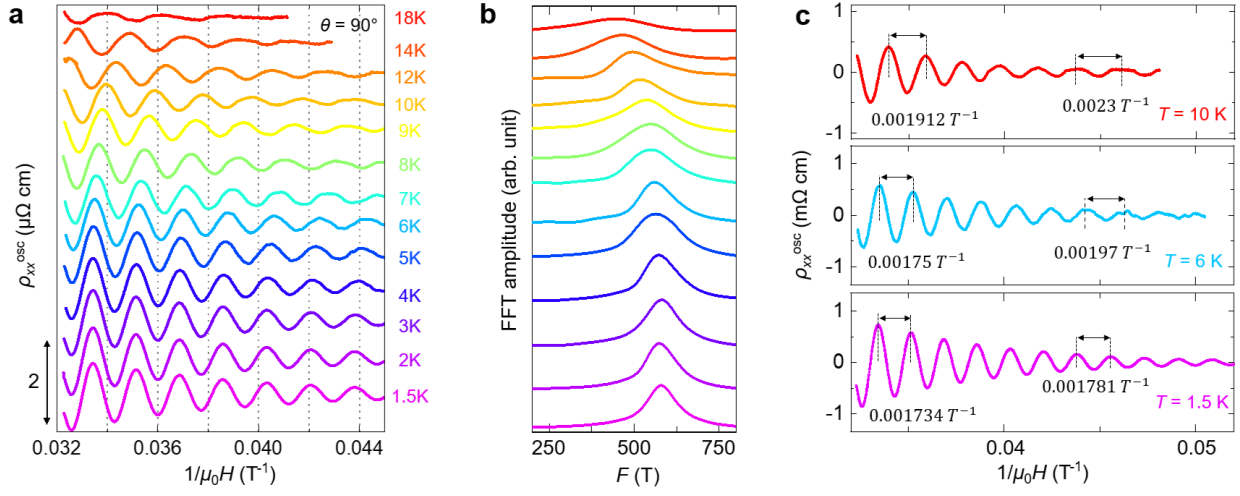

**Fig. S8 | Temperature- and magnetic field-dependences for quantum oscillation. a,** The oscillatory components of SdH oscillations at various temperatures, measured with field along the in-plane direction ($\theta = 90°$), and used to construct the contour plot in Fig. 5g. Data are shifted for clarity. **b**, Fast Fourier transform (FFT) for the oscillatory components in **a**, from which the temperature dependent oscillation frequency shown in Fig. 5h is obtained. Data are shifted for clarity. **c**, Oscillation patterns at 1.5, 6, and 10 K extracted from **a**. The distance between the neighboring oscillation peaks, which is the SdH oscillation period, changes with magnetic field. The field-dependent oscillation frequency presented in Fig. 5i is obtained from inversing the period (see Methods).

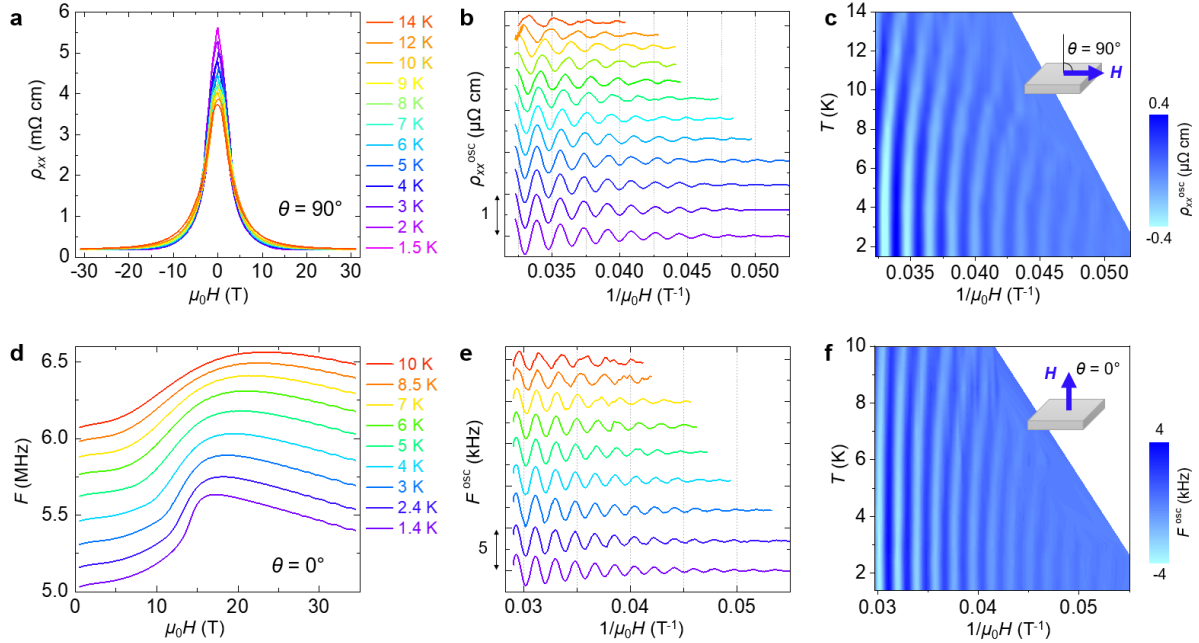

**Fig. S9 | Temperature- and field-dependences for SdH and dHvA effects in additional samples. a,** Field dependence for in-plane resistivity at various temperatures for a sample different from that in Fig. 5f-j, measured with field along the in-plane direction ($\theta = 90°$). The corresponding SdH oscillation components are shown in **b**. **c**, Contour plot showing the temperature and field dependencies for SdH effect, constructed using data in **d**. Field dependence for TDO frequency at various temperatures, measured with field along the out-of-plane ($\theta = 0°$) direction, The corresponding dHvA oscillation components are shown in **e**. **f**, Contour plot showing the temperature and field dependencies for dHvA effect, constructed using data in **e**.

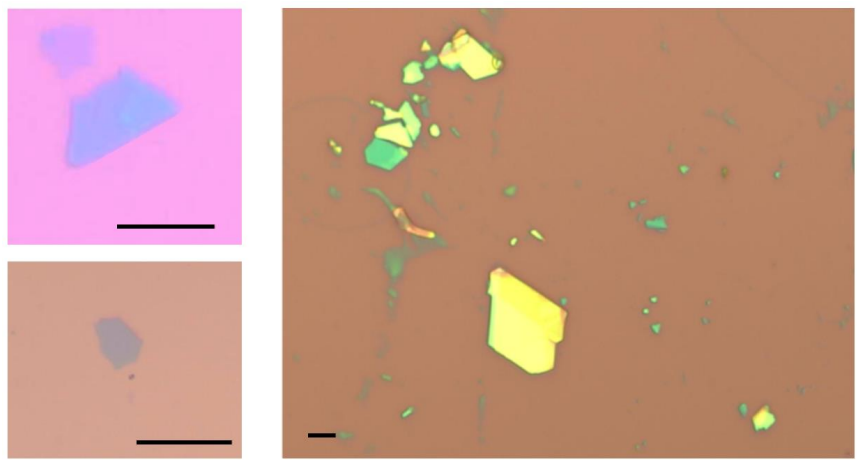

**Fig. S10 | GdPS flakes.** Exfoliated GdPS thin flakes on Si wafer. Scale bar: 10 μm.

## Crystal growth and phase characterizations

Single crystal growth for GdPS is described in Methods. Fig. S1 shows the single crystal XRD and images of the crystals. The crystal structure information obtained from single crystal XRD is provided below in Table S3.

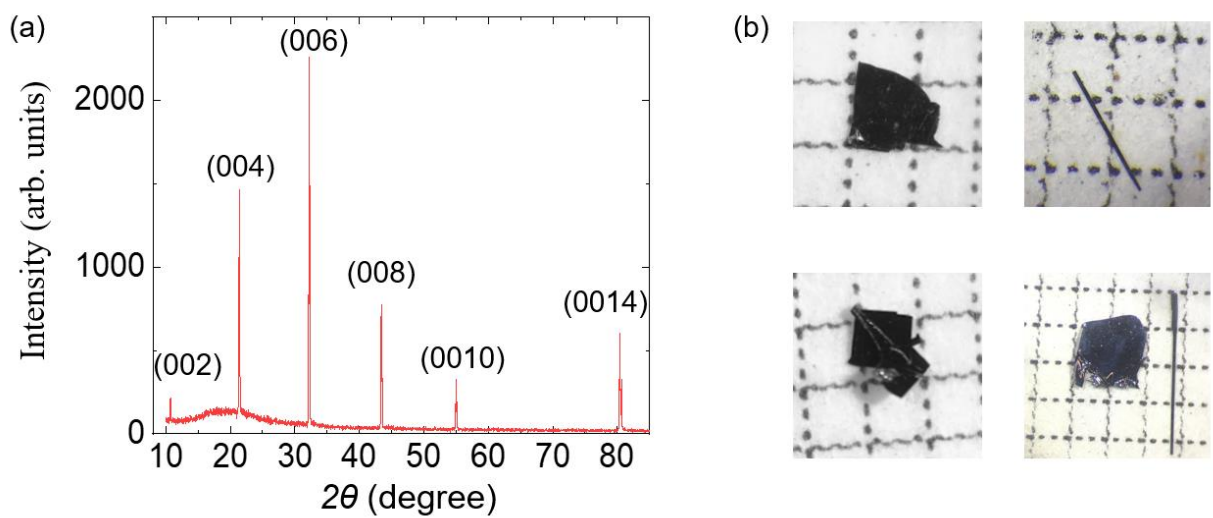

**Fig. S11 | Single crystal growth for GdPS. a,** (00$L$) x-ray diffraction peaks for single crystalline GdPS. **b,** Images of GdPS single crystals on 1mm grid graph papers, showing both needle-like and plate-like crystals.

**Table S3 | Structure parameters for GdPS obtained from single crystal XRD refinement.**

Space group: *Pmnb*

| Lattice constants (Å) | | | | | | Atomic Positions | | | | |
| --- | --- | --- | --- | --- | --- | --- | --- | --- | --- | --- |
| *a* | *b* | *c* | *α* | *β* | *γ* | Atoms | *x* | *y* | *z* | Occupancy |
| 5.36180 | 5.41350 | 16.75150 | 90° | 90° | 90° | Gd1 | 0.25000 | 0.51490 | 0.35362 | 1 |
| | | | | | | Gd2 | -0.25000 | 0.01560 | 0.36263 | 1 |
| | | | | | | P | 0.03830 | 0.29750 | 0.49760 | 0.95 |
| | | | | | | S1 | 0.25000 | 1.01220 | 0.31770 | 1 |
| | | | | | | S2 | -0.25000 | 0.51330 | 0.31180 | 1 |

## Determine the ground state magnetic structure for GdPS

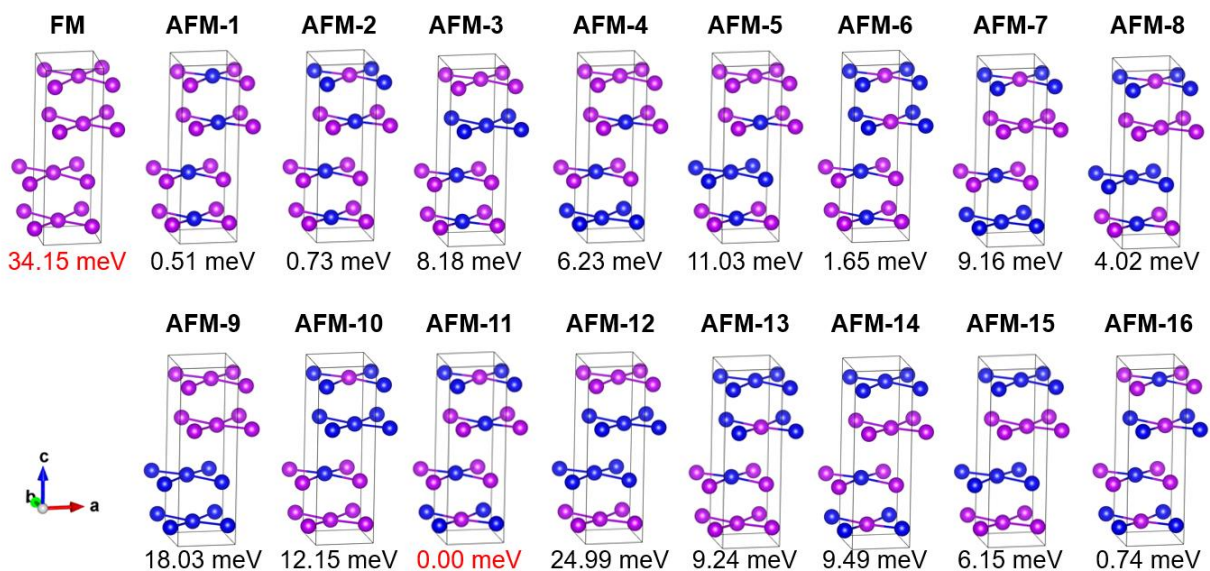

**Fig. S12** | Energy calculations for various possible magnetic structures for GdPS. Only Gd atoms are shown. Purple and the blue color denote opposite spin orientations for AFM state.

According to the structural symmetry, a total of 16 magnetic configurations can be obtained for the antiferromagnetic states (AFM), which are labeled by AFM-1 to AFM-16 in above Fig. S12. The configuration of AFM-11 phase has a lowest energy, and the relative energy is calculated by the formula, $E_R = E_{model} - E_{AFM-11}$, and shown in the bottom of each configuration. The ferromagnetic (FM) state is also shown for comparison.

The predicted AFM ground state is consistent with experimental observation of the lack of irreversibility (Figs. S13a-b) and the low field linear field-dependent magnetization (Fig. 3a, main text). The comparison between out-of-plane and in-plane magnetic susceptibility reveals tiny anisotropy (Fig. S13c).

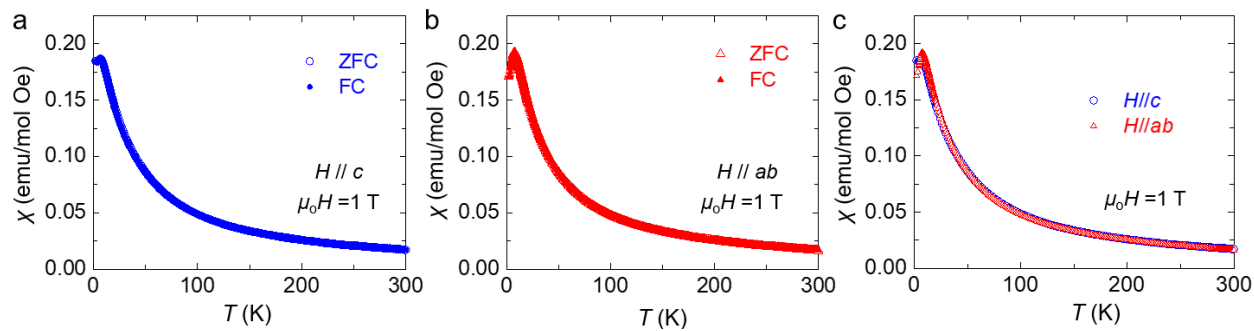

**Fig. S13** | **a-b**, Zero-field-cooling (ZFC) and field-cooling (FC) temperature dependent magnetic susceptibility for GdPS measured with (**a**) out-of-plane (blue, $H//c$) and (**b**) in-plane (red, $H//ab$) magnetic fields of 1 T. **c**, Comparison between out-of-plane and in-plane magnetic susceptibility.

## Calculation for MAE

Table S4 shows the MAE in spherical polar coordinates ($r$, $\theta$, $\varphi$), where $r$, $\theta$, and $\varphi$ represent the radial distance (in our case, the magnetic moment), polar angle, and azimuthal angle, respectively. We observed that in the case of AFM GdPS, the configuration with the magnetic moment oriented at $\theta = 45°$ and $\varphi = 45°$ is slightly more stable than other AFM configurations. Similarly, in the case of FM GdPS, the configuration with the magnetic moment oriented at $\theta = 0°$ and $\varphi = 0°$ is slightly more stable than other FM configurations. We consider these stable structures as references for MAE calculations.

**Table S4** - The variation of MAE with polar angle $\theta$ and azimuthal angle $\varphi$ in FM and AFM GdPS.

| $\theta$ (degree) | $\varphi$ (degree) | MAE for the AFM state (μeV/p.c.) | MAE for the FM State (μeV/p.c) |
|---|---|---|---|
| 0 | 0 | 11.16 | 0.00 |
| 45 | 0 | 9.06 | 0.65 |
| 45 | 45 | 0.00 | 1.13 |
| 45 | 90 | 9.83 | 1.23 |
| 45 | 135 | 22.08 | 1.19 |
| 45 | 180 | 26.44 | 0.72 |
| 45 | 225 | 23.70 | 1.08 |
| 45 | 270 | 25.16 | 1.28 |
| 45 | 315 | 21.10 | 1.15 |
| 90 | 0 | 17.97 | 0.15 |
| 90 | 45 | 11.03 | 1.49 |
| 90 | 90 | 19.33 | 1.49 |
| 90 | 135 | 25.56 | 1.58 |
| 90 | 180 | 17.48 | 0.10 |
| 90 | 225 | 10.48 | 1.51 |
| 90 | 270 | 18.93 | 1.50 |
| 90 | 315 | 25.29 | 1.48 |